\documentclass{phbauth}        
\usepackage{graphicx}

\begin{document}

\begin{frontmatter}

\title{Spectral Features of the Proximity Effect}

\author[address1]{S. Pilgram},
\author[address2]{W. Belzig},
\author[address1]{C. Bruder\thanksref{thank1}}

\address[address1]{Departement Physik und Astronomie, Universit\"at Basel,
Klingelbergstr. 82, CH-4056 Basel, Switzerland}
\address[address2]{Theoretical Physics Group, Delft University of Technology,
2600 GA Delft, The Netherlands}

\thanks[thank1]{Corresponding author. E-mail: bruder@ubaclu.unibas.ch} 

\begin{abstract}
We calculate the local density of states (LDOS) of a
superconductor-normal metal sandwich at arbitrary impurity
concentration. The presence of the superconductor induces a gap in the
normal metal spectrum that is proportional to the inverse of the
elastic mean free path $l$ for rather clean systems. For a mean free
path much shorter than the thickness of the normal metal, we find a
gap size proportional to $l$ that approaches the behavior predicted by
the Usadel equation (diffusive limit).
\end{abstract}


\end{frontmatter}

A superconductor that is in good metallic contact to a normal metal
induces a finite pair amplitude on the normal side. As a result, the
normal metal acquires superconducting properties like infinite
conductance and the Meissner effect (see \cite{superlattices} and
references therein). The presence of the
superconductor also has consequences on the spectrum: the constant
density of states around the Fermi energy of a bulk normal metal 
is strongly modified and shows a pseudogap for a clean normal
metal. That is, it vanishes at the Fermi energy and rises linearly close
to it \cite{DOSLIN}. Another well-known result on the spectrum has
been obtained in the dirty (diffusive) limit
\cite{golubov,belzigdos}: in this case, there is a gap in the spectrum
(called {\it minigap} to avoid confusion with the gap of the
superconductor) that has the same order of magnitude as the Thouless energy
$E_{Th}=\hbar D/d^2$, here $D$ is the diffusion constant of the normal
metal and $d$ its thickness.

In related work, the spectrum of a ballistic normal cavity connected
to a superconductor has been studied. A classically integrable cavity shows
a LDOS that is linear in energy, whereas a chaotic cavity
exhibits a gap in the spectrum \cite{chaos}. 

\begin{figure}[t]
\begin{center}\leavevmode
\includegraphics[width=0.8\linewidth]{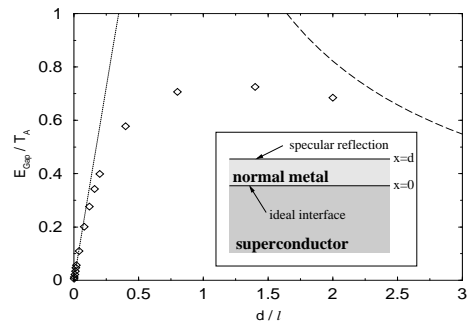}
\caption{Size of the minigap. Diamonds: numerical results. Dotted line:
estimate in the almost clean case as given in the text.  
Dashed line: numerical fit to the dirty-limit results obtained in
\protect\cite{belzigdos}. Inset: geometry of the SN-junction.  
}\label{gapsize}\end{center}\end{figure}

The goal of the present work is to investigate a (moderately)
disordered normal metal and to study in detail how the linear rise of
the LDOS for the clean system transforms into the minigap in the
diffusive system. To this end, we have solved the real-time
Eilenberger equation for the quasiclassical $2\times 2$ matrix Green's
function $\hat g$ numerically for the planar geometry shown in the
inset of Fig.~\ref{gapsize} (see \cite{superlattices} for additional
details of this method). Impurity scattering is taken into account by
including an impurity self-energy of the form $\langle \hat g
\rangle/2\tau$ (Born approximation) determined in a self-consistent
way. The SN-interface is assumed to be ideal, and the outer boundary
of the normal metal is specularly reflecting. The normal metal is
characterized by a vanishing pairing interaction.
We have neglected
self-consistency of the pair potential $\Delta(x)$ in the
superconductor. The slight suppression of $\Delta(x)$ due to the proximity
effect does not affect the low-energy spectrum of the normal metal since we
assume $d\gg \xi_0=\hbar v_F/\Delta$.

\begin{figure}[t]
\begin{center}\leavevmode
\includegraphics[width=0.8\linewidth]{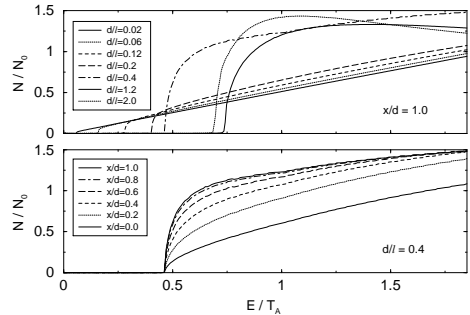}
\caption{Local density of states of the normal side of a proximity sandwich. 
$d$ is the normal-layer thickness and $l$ the elastic
mean free path (both in N and in S);  $T_A=\hbar v_F/2\pi d$); $d=1000\xi_0$.
Upper panel: Adding impurities leads to the formation of a minigap.
Lower panel: The minigap is constant throughout the normal metal, but the 
energy dependence of the LDOS changes with location.
}\label{ldos}\end{center}\end{figure}

As a result of a detailed numerical study, we find that a gap forms at
arbitrarily small impurity concentrations. This is shown in the upper
panel of Fig.~\ref{ldos}: even for values of the elastic mean free path $l$
that are $50$ times larger than the normal-layer thickness, the
formation of the low-energy gap is clearly visible. The gap increases
with $1/l$, saturates for $l\sim d$ and then decreases again as
expected from the dirty-limit theory since $D\sim l$. The gap does not
depend on the location in the normal metal as can be seen in the lower
panel of Fig.~\ref{ldos}, i.e., it is a global feature. The shape of
the LDOS, i.e., its dependence on energy, however, varies on
traversing the normal layer.

Figure~\ref{gapsize} shows the dependence of the minigap on $d/l$. The
behavior can be understood in a qualitative way as follows: the linear
dependence of the LDOS on energy in the clean case is caused by the
existence of Andreev bound states with energies $E_n(\theta)=\pi \hbar
v_F (n+1/2) \cos \theta /2 d$ with $\theta\sim \pi/2$. Here, $\theta$ denotes
the angle of a semiclassical trajectory with the normal vector of the
interface. These trajectories (V-shaped structures bounded by 
two Andreev reflections and one specular reflection at the outer interface) 
have a
length $\sim d/\cos \theta$. In almost clean systems, this length is
cut off by the mean free path, hence there is a lower energy scale of
$E_{Gap}\sim \hbar v_F/l\sim T_A d/l$ where $T_A=\hbar v_F/2\pi d$ is
the Andreev temperature. The appropriate energy scale for the diffusive limit
can be obtained as $\hbar/t$ where $t$ is defined by $d \sim \sqrt{Dt}$.
As a result, we obtain $E_{Gap} \sim \hbar D/d^2 \sim T_A l/d$.

In conclusion, we have solved the real-time Eilenberger equation
numerically and have determined the local density of states of a
proximity sandwich. We find a hard gap that opens for an arbitrarily
small concentration of impurities and is maximal if the elastic mean
free path is of the order of the normal-layer thickness.



\end{document}